\documentclass[aps,prb,twocolumn,showpacs,superscriptaddress,floatfix,citeautoscript]{revtex4-1}
\usepackage{times}
\usepackage{graphicx}
\usepackage{dcolumn}
\usepackage{bm}
\usepackage{color}
\begin{document}

\title{Melting Scenario for Coulomb-interacting Classical Particles in Two-dimensional Irregular Confinements}

\author{Dyuti Bhattacharya}
\affiliation{Indian Institute of Science Education and Research-Kolkata, Mohanpur Campus, India-741252}
\author{Amit Ghosal}
\affiliation{Indian Institute of Science Education and Research-Kolkata, Mohanpur Campus, India-741252}

\begin{abstract}
The ``melting" of self-formed rigid structures made of a small number of interacting classical particles confined in an irregular two-dimensional space is investigated using Monte Carlo simulations. It is shown that the interplay of long-range Coulomb repulsions between these particles and the irregular confinement yields a solid-like phase at low temperatures that possesses a bond-orientational order, however, the positional order is depleted even at the lowest temperatures. Upon including thermal fluctuations, this solid-like phase smoothly crosses over to a liquid-like phase by destroying the bond-orientational order. The collapse of solidity is shown to be defect mediated, and aided by the proliferation of free disclinations. The behavior of different physical observables across the crossover region are obtained. Our results will help quantifying melting found in experiments on systems with confined geometries. 

\end{abstract}

\maketitle

\section{Introduction}

The phenomenon of melting \cite{HuangBook63,Book65,Ramakrishnan79,Frenkel80,RMP2D88} has always fascinated  physicists from the
early dawn of study of the condensed phases, because it is encountered in everyday life. The order (or quasi-order) characterizing a solid, as well as the mechanism
through which a solid melts, depends crucially on the spatial dimensionality
\cite{Dimen01}. For example, a two-dimensional (2D) system presents additional challenges
in comprehending melting than three-dimensional one because of the increased significance of
fluctuations \cite{MWth66}. While comprehensive research effort has enriched a coherent
understanding of 2D melting in bulk systems \cite{RMP2D88}, the
physics of melting in confined systems made of small number of particles has not been studied as much. One cannot
expect a sharp phase-transition \cite{YangLee52} in a confined
system unlike in the bulk -- a thermodynamic phase itself is ill-defined
with a finite number of particles. However, some signatures of the ``melting'', say, a Crossover (CO) from a solid-like to a liquid-like phase
have been found \cite{GhosalNat07,Expt1stpara11}.

Such 2D-clusters find significant experimental relevance and are realized
in a variety of experiments, such as, radio-frequency
ion traps \cite{Expt609}, electrons on the surface of liquid $\rm{He}$
\cite{Expt412}, electrons in quantum dots in semiconductor heterostructure
\cite{QD07}, and in dusty plasmas \cite{ExptRev10}. The interactions between the constituent charged particles are expected to maintain the bare long range ($\sim r^{-1}$) Coulomb repulsion due to the lack of screening in small clusters compared to the bulk. The resulting solid-like
phase in a cluster has been termed as a Wigner Molecule (WM),\cite{QD07}
due to its analogy with the Wigner Crystal (WC) in bulk
systems \cite{2DWC34}. Incidentally, a 2D WC undergoes melting
to a Fermi liquid even at zero temperature ($T=0$) \cite{Tanatar89},
driven by quantum fluctuations, and so does a WM in circularly
symmetric traps \cite{Fillinov01,Ghosal07,Guclu08}. However, for a variety
of experimental clusters at low $T$, the quantum zero-point motion, inherent to particles in a bound states, and the associated quantum fluctuations are not so important: either due to the finite operating $T$, or because of the heavy mass of the particles
or their low density, or a combination of various experimental reasons. The rich
physics of the CO is thus contributed entirely by the classical thermal
fluctuations. Temperature driven melting of classical Coulomb-interacting
2D plasma was studied numerically \cite{GannSudip79} in the past,
and the corresponding Kosterlitz-Thouless-Halperin-Nelson-Young (KTHNY)
theory of 2D melting has also been developed. \cite{KT73,Halperin78,Young79,Nelson79}
Similar thermal melting of clusters in 2D harmonic trap
has recently been studied extensively \cite{BedPeet94,Fillinov01}, and
also been realized experimentally \cite{Plasma12}, leading to a detailed insight
through the measurements of various observables including relative positional fluctuations
and associated Lindemann ratio \cite{Plasma12}, pair-correlation function
\cite{phTrans_finiteSys96}, static structure factor \cite{sofk06}
and addition spectra.\cite{addSp99}

Progresses in the field of high accuracy fabrication made the clusters ultra-tunable and the shape of confinement is controlled at will by electrostatic and magnetic
methods \cite{ExptRev10}. These systems naturally have become a hotbed
for systematic study of a complex interplay of Coulomb-repulsion, quantum
interference effects of the confinement, level quantization due
to their smallness, and finally, disorder in the form of irregularities
in the confinement. Existence of irregularities, at least
in large quantum dots has been established in Coulomb blockade experiments
\cite{Marcus97,Sivan96}, and corresponding theories based on ``Universal Hamiltonian''
\cite{Alhassid00,Glazman06} have been put forward. It is this last point
-- the effect of disorder or irregularities on the melting mechanism, 
that has drawn a significant attention in the recent past \cite{RMP01}
in the context of the quantum melting in 2D WC.\cite{chuitanatar95} While the melting scenario is still unsettled even for the disordered bulk 2D systems, with proposals of intermediate exotic
phases \cite{Waintal06,spivak10}, the role of irregularities
on the melting of WM has received relatively little attention (See however, \onlinecite{Koulakov96,yuval99,peeters06,egger03}) and is addressed
in this work in details.

There are relevant questions with respect to the CO from a solid-like to a liquid-like phase in an Irregular Wigner Molecule (IWM): Is there a low-temperature solid-like phase at all, in spite of the irregularities? Does the CO occur more or less abruptly or gradually with $T$? What are the reasonable criteria for identifying the possible CO in an IWM? And finally, what is the physical mechanism governing the CO, if any?
While addressing these important questions within a framework of a systematic calculation, we
first show that there indeed exists a phase that is solid-like, where the solidity is arising primarily from the orientational order at low $T$.
Upon inclusion of thermal fluctuations, the IWM crosses over to a liquid-like
phase. From a detailed study of the behavior of different observables, we
estimate the temperature width $\Delta T_{X}$ for the crossover. We also present compelling  numerical evidences illuminating the mechanism for this crossover, which turns out to be associated with formation and proliferation of free disclinations on top of the irregular geometry that diminishes the positional order in the IWM.

\section{Model, Parameters and Method}\label{sec:model}

We consider a system of $N$ classical particles interacting
via a Coulomb repulsion $V_{\rm Coul}=\sum_{i<j}\left(\vec{r}_{i}-\vec{r}_{j}\right)^{-1}$ (The standard Coulomb energy factor $C=q^2/\varepsilon$ is assumed to be unity). The effect of irregularity is introduced through the following 2D irregular quartic confinement potential $V_c$:
\begin{equation}
V_c(x,y)=a\left[\frac{x^{4}}{b}+by^{4}-2\lambda x^{2}y^{2}+\gamma(x-y)xy r\right],
\end{equation}
where $r=\sqrt{x^{2}+y^{2}}$. The parameter $a$ makes the confinement narrow or shallow and thus controls the average density of particles in the system. We present all our results for $a=1$. Effect of a change in the value of $a$ will be commented on later in this report. We choose $b=\pi/4$ \cite{Bohigas93} that breaks $x$-$y$ symmetry of the quartic oscillator. $\lambda$ in Eq. (1) controls the chaoticity, and $\gamma$ breaks the reflection symmetry.

Our interests lie primarily in unfolding the universal features of disordered systems, it is thus essential that the chosen $V_c$ respects those. Fortunately, there exists a large body of literature confirming the universal behavior of chaotic quantum dots with our choice of $V_c$ in Eq. (1). This is particularly important because a `soft' potential (with smooth boundary like ours, and relevant for experiments where confinement is set up by electrostatic and magnetic means) rarely shows a wide range of chaotic behavior, unlike the billiards \cite{Ullmo03}. The classical dynamics of non-interacting particles in the above confinement has been studied in details \cite{Bohigas93,Ullmo03} with the identification of the integrable and the chaotic regimes. The interplay of Coulomb-repulsion between spin-$1/2$ fermions and the irregularities of $V_c$ have also been
looked at \cite{hong03,ghosalhong05}. With the aim of generating data on self-similar copies of irregularities for better statistics, we choose $\lambda\in[0.565,0.635]$
and $\gamma\in[0.1,0.2]$ and generate $5$ different realizations of confinement, each for a given combination of $(\lambda,\gamma)$. Finally, we study the behavior of $N=10$ to $151$ particles and the thermal evolution of several observables described in the next section.

While the particles have no dynamics at $T=0$ due to the lack of zero-point motion, many low-energy configurations with updated location of particles contribute to the partition function for $T>0$. As a result, the thermal contribution to the physical observables is weighted with appropriate Boltzmann factors. These observables thus carry in their thermal evolution the complex interplay of thermal kicks on these classical particles and their inter-particle Coulomb potential energy, as well as the effect of the external irregular confinement. The thermal evolution of the system is studied
using standard Monte Carlo (MC) simulated annealing \cite{cerny85}
aided by Metropolis algorithm \cite{metro53}. Simulated annealing is expected 
to track the appropriate low-energy states consistent with the Boltzmann probability
at a given $T$ ($T$ includes $k_B$, the Boltzmann factor for all our results), and can obtain the Ground state configuration in the limit $T \rightarrow 0$,
for an appropriate choice of a slow annealing schedule \cite{kirk83}.

\section{Results}\label{sec:res}

We describe our results by showing evidences for the `solidity' in the proposed IWM at $T=0$. An example of a ground-state configuration (GSC) is shown in Fig. 1(a) for a fixed set of parameters (See the caption). Such a spatial configuration of particles (for all different realizations) was obtained by running the simulated annealing to a very low $T\sim10^{-5}$, and then letting the system relax to $T=0$ configuration following the downhill moves alone in the energy landscape. While it is not possible to access the true GSC for each individual run, we ensure that the GSC is indeed achieved by the following statistical analysis: We start the $T=0$ calculation from $\sim200$ different configurations generated
in the equilibrium MC runs at the lowest $T$, for a given realization. Starting with these as initial configurations, not all of the $200$ runs converge at $T=0$ to the
same final GSC. However, for all the cases we study in the entire parameter space, more than $60\%$ of them do, with exactly the same energy $E_{0}$. Furthermore, the runs for which the final ground-state at $T=0$ are different from the true GSC (with energy $E_{0}$)
always had an energy, $E > E_{0}$, giving us the confidence of identifying
the true GSC. We also ensure that all the forces (originating from the inter-particle potential and confinement) on each particle in the true GSC add up to zero, modulo a numerical tolerance, by using first principle Newtonian mechanics.

\begin{figure}[t]
\includegraphics[width=3.3in,height=3.3in]{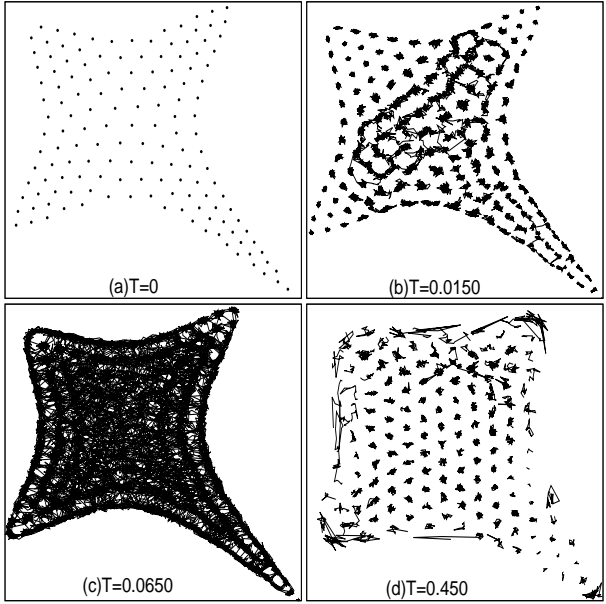}
\caption{(a) The GSC (at $T=0$) with $N=148$ particles in one realization of the confinement $V_c$ of Eq. (1) with $\lambda=0.2$, $\gamma=0.635$, showing the amorphous solid-like behavior. Snapshots of particles in the same confinement are shown in (b) and (c) at larger temperatures. (b) At $T=0.015$, where the melting has just started, the spatially correlated `movements' of several particles in the MC configuration space have produced a tortuous path connecting these particles. (c) At $T=0.065$ the melting is nearly complete. $100$ independent MC configurations were used for generating the snapshots, and these MC configurations were taken from the equilibrium MC runs separated by the intervals in which the particles would have moved the same distance diffusively. For comparison, similar snapshot is presented in (d) for hard-core particles (core size $\sim 0.285$) in the same confinement, showing the `pre-melting' primarily occurring near the boundaries.}
\label{fig:config}
\end{figure}

Snapshots of  the MC configurations, tracing the paths traversed by individual particles in space during MC evolution, are presented in Fig. 1(b) and (c) for $T=0.015$ and
$0.065$ respectively, in the unit of energy. These snapshots play important role in visualizing the crossover
from a solid-like to a liquid-like behavior in the IWM in the following manner: In Fig. 1(b), where the ``melting" has just started, the thermal motion
of some particles becomes correlated leading to the incipient melting through the
formation of a tortuous path. While such paths (for a given $N$ and realization) always occur at similar $T$, signaling the commencement of melting, their locations are completely random in space without
any preference to the bulk or to the boundary. We will bring back our attention to
these interesting structures when we discuss the possible mechanism of
the crossover. Fig. 1(c) illustrates the ``melted" state,
where every particle travels almost everywhere in the system.  
Fig. 1(d) presents a similar snapshot in the same confinement depicting the incipient melting for hard-core particles \cite{hardcore00},
with the average core size same as the Coulomb-interacting particles in the IWM as inferred from the pair distribution function (defined below).
Comparison of Fig. 1(b) and Fig. 1(d) shows that the melting starts predominantly on the boundary for the hard-core particles, unlike the Coulomb-interacting ones. While pre-melting on the boundary in solids with short-range interactions has been discussed for a long time \cite{surfacemelting88,surfacemelting89}, our results with long-range Coulomb repulsion show that it can occur anywhere in the system, based on the statistics  from different realizations of the confinement.

Having established that we indeed track the true GSC, the next question
is: Are these IWM equivalent to the solid-like phase in confined
systems? Even the bulk 2D WC leads to very interesting physics.\cite{goldman90,Tanatar89} Extensive research \cite{spivak05}
confirmed that an ideal 2D WC phase is characterized by both positional
and orientational orders, and the collapse of both are necessary to
cause its final melting, and thus it is likely to occur in two stages \cite{RMP2D88,Young79}. Fig. 1(a) clearly shows that the Positional Order (PO) is already
depleted even at $T=0$, because of the breaking of translational symmetry
by the confinement, yielding an amorphous solid-like phase. Lack of positional order is also verified by calculating $\rho_{\vec{k}}=N^{-1}\sum_{i}\exp(i\vec{k}.\vec{r}_{i})$, $\vec{r}_{i}$ being position of $i$-th particle. $\rho_{\vec{k}}$ shows only broad humps for IWM even at $T=0$ instead of sharp Bragg-peaks. This points towards the absence of PO down to $T=0$. We do not see any significant thermal evolution of $\rho_{\vec{k}}$ either.

The orientational order, on the other hand, is evident from the nearly perfect $6$-coordinated environment for all the particles, except obviously for those on the
boundary. We reiterate that the orientational order, together with the positional order, characterizes solidity in the 2D bulk systems. Our results also indicate that a self-formed IWM makes approximately a triangular lattice (modulo the irregular lattice lines in the absence of PO) -- typical 2D Bravais lattice minimizing the total energy with long-range interactions. A quantitative estimate of the orientational order is measured through Bond-Orientational Order (BOO) \cite{Nelson79},
$\psi_{6}(\vec{r})$ (See Fig. 3), and the Bond-Orientational Correlation Function (BOCF), $g_{6}(r)$, to be defined in the next section (See Fig. 4).

The other signature of solidity, even for the amorphous solid-like phase in an IWM,
lies in its rigidity prohibiting any significant root mean square fluctuations of the constituent particles from their equilibrium positions at low
$T$. Such fluctuations measured in terms of the average inter-particle
spacing are known as Lindemann Ratio (LR), ${\cal L}$ \cite{Linde10}. Our results in Fig. 2 demonstrate that the average LR is essentially zero for all $T<0.01$, and thus provide further evidence for the solid-like IWM at $T=0$.

\begin{figure}[t]
\includegraphics[width=3.3in,height=3.2in]{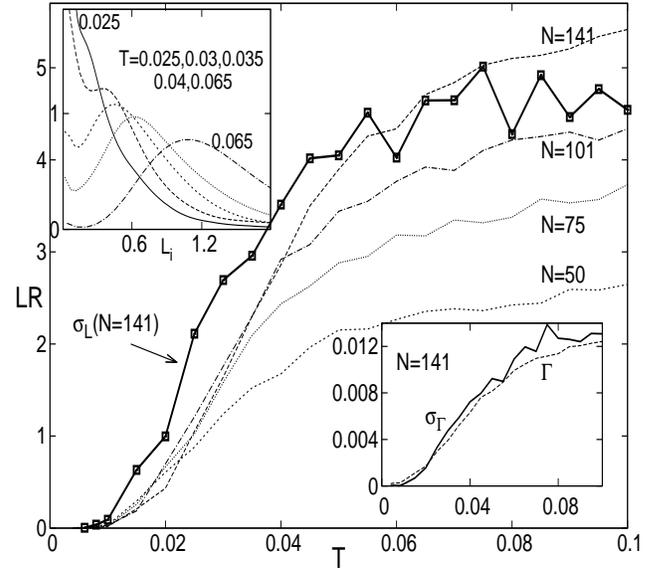}
\caption{The main panel shows the evolution of ${\cal L}$ (defined in the text) with $T$ for different $N$. Each trace for a given $N$ is averaged over $5$ realizations of confinement. While ${\cal L}(T<0.01) \approx 0$, its smooth rise in the range of $0.01 \leq T \leq 0.05$ indicates 'melting'. For $T \geq 0.06$, ${\cal L}$ still rises but with a much weaker slope, as expected from diffusive motion of particles in a `fluid' phase. The top-left inset presents the $T$-evolution of the probability distribution of ${\cal L}_i$, $P({\cal L}_i)$ for $N=141$. The variance of $P({\cal L}_i)$, $\sigma_{\cal L}$, (scaled by a factor $155$ for clarity) is also presented in the main panel, showing similar rise as ${\cal L}$ itself. The bottom-right inset describes the same crossover as in main panel, but in terms of modified Lindemann parameter $\Gamma$ (defined in the text), and its variance $\sigma_{\Gamma}$ (scaled by $1200$) for $N=141$.
}
\label{fig:lindeman}
\end{figure}

How does an IWM melt with increasing $T$ ? Insight could be derived
from the LR given as: ${\cal L}=N^{-1}\sum_{i}{\cal L}_{i}$,
where ${\cal L}_{i}=a_{i}^{-1}\sqrt{\langle | \vec{u}_i |^2 \rangle}$
for $i$-th particle, and $\vec{u}_i=\vec{r}_{i}-\vec{r}_{i}^{(0)}$. Here $\vec{r}_{i}^{(0)}$ is the position of the $i$-th particle in the initial configuration (i.e., the one at the end of the equilibration MC steps), and $a_i$ is the average distance of the $i$-th particle with its neighbors. We track the thermal evolution of ${\cal L}$ (averaged over all realizations)
in Fig. 2. While ${\cal L}(T<0.01)\approx 0$, LR increases dramatically
for $0.01\leq T\leq0.05$, beyond which the growth of ${\cal L}$
becomes far more gradual. This range of temperature 
thus identifies the crossover width, $\Delta T_{X}$, between the IWM
and its melted state. Distribution of ${\cal L}_{i}$, $P({\cal L}_i)$, collected over all particles and also over realizations of confinement is shown in the top-left
inset of Fig. 2 for $0.025 \leq T \leq 0.065$. It indicates that a peak in $P({\cal L}_i)$ starts appearing for non-zero values of ${\cal L}_{i}$ for all $T\geq0.03$. The variance, $\sigma_{{\cal L}}$, of $P({\cal L}_i)$ follows a similar $T$-evolution as
the ${\cal L}$ itself, as presented in the main panel of Fig. 2, giving further
confidence in identifying $\Delta T_{X}$ as the crossover width.

The above definition of ${\cal L}$ is difficult to implement for the bulk 2D systems, because, $\langle | \vec{u} |^2 \rangle$ can show logarithmic divergence \cite{LandauPeierls} with system size. Therefore a modified Lindemann parameter, $\Gamma$, has been proposed \cite{Bedanov85} in terms of the relative inter-particle distance fluctuations:
\begin{equation}
\Gamma = \left \langle \frac {1}{N} \sum_{i=1}^N \frac{1}{a_i^2 N_b}\sum_{j=1}^{N_b} \left ( \vec{u}_i - \vec{u}_j \right )^2 \right \rangle^{1/2},
\end{equation}
where, the summation on $j$ runs over $N_b$ number of nearest neighbors of $i$-th particle. $\Gamma$ is free from any divergence and it has been used extensively to track 2D melting \cite{BedanovJETP,Hartman07}. However, particles are confined in an IWM within a finite length-scale, and thus, even ${\cal L}$ remains free of divergence in an IWM in contrast to the bulk. We, nevertheless, calculated the evolution of $\Gamma$ and the variance $\sigma_{\Gamma}$ of $P(\Gamma_i)$, defined exactly in the same way as those for ${\cal L}_i$. The resulting $\Gamma$ and $\sigma_{\Gamma}$ are shown as a function of $T$ in the bottom-right inset of Fig. 2 for $N=141$. The similarity of their $T$-evolution with those for ${\cal L}$ serves as a consistency check for our results.

While the thermal evolution of $\Gamma$ shows a sharp transition (i.e., $\Delta T_{X} \rightarrow 0$) for the bulk 2D system \cite{Bedanov85},a confined (and hence finite) system would always have a finite $\Delta T_{X}$
We found that its value for IWM is typically larger than that for a circularly symmetric WM \cite{BedPeet94}. This can be crudely taken as the smearing of $\Delta T_{X}$ by irregularities.
Our study also indicates that $\Delta T_{X}$ does not change much 
for $N>35$, and does not seem to depend on the model parameters $[\gamma,\lambda]$.
Beyond $T>0.05$ the thermal evolution of ${\cal L}$ follows a weaker linear
trend, as expected for a liquid with particles moving around
diffusively.\cite{Cussler09}  $T$-dependence of LR had been used to identify melting in the bulk (respecting translational symmetry in all directions) \cite{Bedanov85}, as well as in circular dots \cite{BedPeet94,Fillinov01}.In the latter case, the translational symmetry in the azimuthal direction and the lack of it in the radial direction helps defining the radial and the azimuthal LR separately, however, we don't see such separation because our confinement breaks all symmetries.

\begin{figure}[t]
\includegraphics[width=3.3in,height=3.0in]{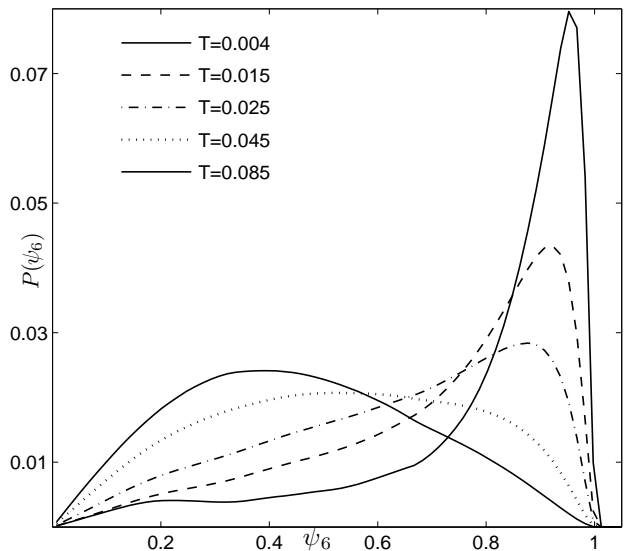}
\caption{The distribution, $P(\psi_6)$, for $N=125$ shows a strong peak at $\psi_6 \approx 1$ signifying nearly regular 6-coordinated IWM at low $T$. $P(\psi_6)$ broadens with $T$ across $\Delta T_X$, finally producing a very broad distribution ranging between $0$ and $1$ beyond the crossover from solid-like to liquid-like phase. The boundary-particles, not having $6$ neighbors surrounding them, were not considered in $P(\psi_6)$.}
\label{fig:boo}
\end{figure}

As mentioned already, GSC in an IWM seems to possess $6$-coordinated bond-orientational order. It is important that we study the temperature
dependence of the local $6$-fold bond-orientational order parameter
\cite{Nelson83}, defined as:
\begin{equation}
\psi_{6}(\vec{r})=\frac{1}{6}\sum_{k=1}^{6}\exp(6i\theta_{k})
\end{equation}
Here $\vec{r}$ is the position of the particle in question, and $\theta_{k}$
is the angle that the particle at $\vec{r}$ makes with its six closest
neighbors $k$, relative to any arbitrary reference axis. $\psi_{6}(\vec{r})$
measures local orientational order in the IWM based on the principle
that all bonds in a perfect triangular lattice should have the same
$\theta_{k}$, modulo $\pi/3$, implying $\psi_{6}=1$ for all $\vec{r}$.
We present the distribution $P(\psi_{6})$ in Fig. 3, for $N=125$ collecting data from all realizations.
The sharp peak in $P(\psi_{6})$ at $\psi_{6}=1$ demonstrates the
persistence of strong BOO in the GSC of an IWM at low $T$, particularly
for large $N$. In a liquid, on the other hand, lack of BOO forces random $\theta_{k}\in [0,2\pi]$, resulting in a broad distribution of $P(\psi_{6})$ ranging between $0$ and $1$. Our result of $T$-dependence of $P(\psi_{6})$
indicates that while the change from solid-like to liquid-like crossover
takes place continuously with $T$, the most dramatic change in which the remnance of the low-$T$ peak of $P(\psi_{6})$ gets washed out, occur in same range of $\Delta T_{X}$ where ${\cal L}$ and $\sigma_{\cal L}$ changes most rapidly (See Fig. 2). This provides further support for the identification of $\Delta T_{X}$ as the CO width. For clusters
with $N\leq35$, $P(\psi_{6})$ develops qualitatively non-universal features, including multiple peaks for some values of $N$, which prohibits reliable identification of $\Delta T_{X}$. The ratio of the number of particles at the boundary to those in the bulk increases resulting in a fewer data-points for constructing $P(\psi_{6})$. Nevertheless, $6$-coordinated bulk particles can be broadly identified down to $N\sim20$, at the lowest temperature. 
\begin{figure}[t]
\includegraphics[width=3.3in,height=3.0in]{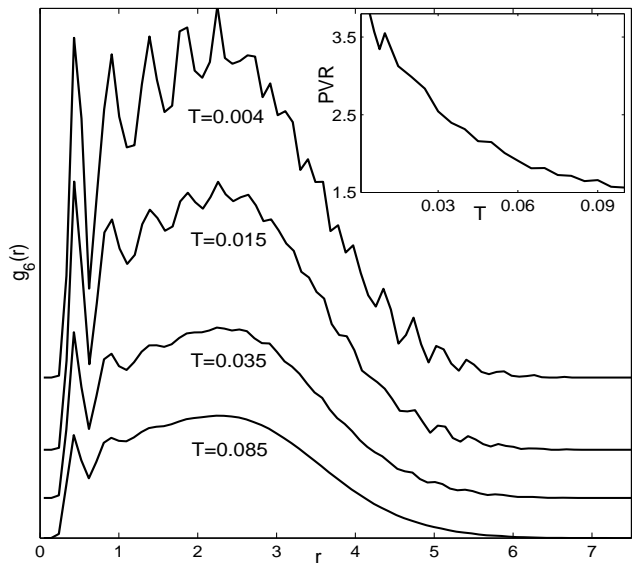}
\caption{Evolution of BOCF, $g_6(r)$ for $N=125$, with $T$, showing clear Bragg-type peaks at low $T$ signifying bond-orientational order. These peaks weaken with increasing $T$, finally producing only couple of initial humps like a liquid. Data for different $T$ are shifted vertically for clarity.
The boundary particles are not taken into account for $g_6(r)$, because they don't have $6$ surrounding neighbors. Inset shows the $T$-evolution of PVR (defined in the text). The smooth decrease of PVR over $\Delta T_X$ is indicative of a smooth CO. }
\label{fig:Gbr}
\end{figure}

While $\psi_{6}(\vec{r})$ describes strength of local BOO, the long-range
nature of this order is best discussed in terms of the associated
bond-orientational correlation function (BOCF), defined as:
\begin{equation}
g_{6}(r) \equiv g_6(|\vec{r}|)=\langle\psi_{6}^{*}(|\vec{r}'|)\psi_{6}(|\vec{r}'-\vec{r}|)\rangle,
\end{equation}
which measures the distance up to which local $\psi_{6}$ are correlated. It is exactly
the same way the pair distribution function (PDF), $g(r)$, defined as:
\begin{equation}
g(r) \equiv g(|\vec{r}|)=\langle\delta(|\vec{r}'|)\delta(|\vec{r}'-\vec{r}|)\rangle,
\end{equation}
which measures the positional correlation, describing the probability of
finding another particle at a distance $r$ from a given one. Our results for BOCF in Fig. 4 at low $T$ show well
defined Bragg-peaks at specific values of $r$ for the largest inter-particle distances in
IWM indicating that the BOO is long-ranged in the scale of linear dimension of the IWM. It is well known that $g_6(r)$ (and $g(r)$ as well) in a bulk system tends to a constant value for $r \rightarrow \infty$ indicating a uniform density. In IWM the distance between any two particles is limited by the linear dimension of IWM and as a result $g_6(r)$ ( and $g(r)$ too) must vanish beyond the system size.
With increasing $T$, the long-range nature diminishes by washing
out the peaks progressively at large $r$, and beyond $T\geq0.05$, only a liquid
like behavior persists featuring only the first couple of humps, as presented in Fig. 4. In the liquid-like phase, $g_6(r)$ (and $g(r)$ as well) is expected to follow the profile of the {\it average} radial density for large $r$, larger than the positions of the initial humps. This is the origin of a smooth $r$-dependent background, we obtain in IWM at large $T$.
The window of $T$, in which the
peaks at large $r$ of $g_6(r)$ starts disappearing, leaving only liquid-like features, is consistent with the $\Delta T_{X}$ reported from the Lindemann analysis.
An obvious quantitative measure of how fast these Bragg peaks are depleting is the peak-to-valley ratio (PVR) -- the ratio of the value of $g_6$ at its first peak at the lowest $r$ to the same at the valley immediately after, is also presented as an inset in the same figure.

\begin{figure}[t]
\includegraphics[width=3.3in,height=3.0in]{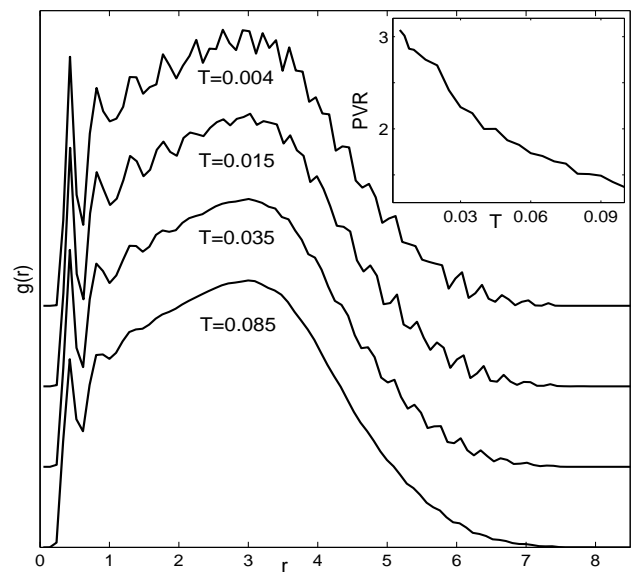}
\caption{Evolution of PDF, $g(r)$ for $N=125$, with $T$, showing much weaker Bragg peaks at low $T$ compared to those in $g_6(r)$ of Fig. (4), indicating weaker positional order than bond-orientational order. The other features follow the same trend of $g_6(r)$. The inset showing a smooth decrease in the PVR of $g(r)$, and this decrease is less drastic than that of $g_6(r)$.}
\label{fig:Gofr}
\end{figure}

We have also studied the thermal evolution of $g(r)$, which is presented
in Fig. 5. The qualitative features are similar to those of $g_{6}(r)$,
except that the peaks in $g(r)$ are less sharply defined. It is known that for 2D bulk system the long-range nature of $g_{6}(r)$
survives up to larger $T$ \cite{RMP2D88} than for $g(r)$, leading to a hexatic
phase \cite{Halperin78} at intermediate temperatures. A reliable identification
of such a phase from numerics \cite{Quinn96} is difficult in the IWM owing to its small linear
dimension. More insights on the CO, however, are obtained from the PDF
of individual particle, $g_{i}(r)$, in the following manner. The 
pair-distribution function of $i$-th particle, $g_{i}(r)$, is defined as the distribution of distances of the $i$-th particle with all the other $(N-1)$
particles averaged over all the MC steps. This definition, in fact, ensures that $g(r)=N^{-1}\sum_i^{N} g_{i}(r)$. 
The nearest neighbor distance of the $i$-th particle, $a_i$, averaged over all the neighbors and also over the MC steps is defined as the value of $r$, where the first peak of $g_{i}(r)$ appears. This same $a_i$ has been used in the Lindemann analysis as the inter-particle spacing of the $i$-th particle. Similarly, the location of the first peak of $g(r)$ defines the mean inter-particle spacing for the system.

While every particle travels through the whole system for large $T$ ensuring $g(r) \approx g_{i}(r)$, such an independence of $g_{i}(r)$ on $i$ does not hold at small $T$. For $T \rightarrow 0$, $g_{i}(r)$ is determined by the environment of the $i$-th particle, which could differ significantly from that of any other particle due to irregular confinement.
Thus, the low-temperature distribution,
$P(\Delta g)$, of $\Delta g(r) \equiv g_{i}(r)-g(r)$ collected for all $r$, and also over the realizations of irregularity, will be rather broad, and need not even be symmetric. The asymmetry in $P(\Delta g)$ for a given realization of $V_c$ depends on its parameters, and is washed out in the ensemble averaging over many realizations. On the contrary, $P(\Delta g)$ is narrow and symmetric at larger $T$, because all particles are equally likely to be everywhere in the system.
It is then expected
that the thermal evolution of $P(\Delta g)$ must shed some light on the crossover.
Note that a similar argument for the bulk system would lead to a symmetric and possibly narrow
$P(\Delta g)$ for both large and small $T$. We present the $T$-dependence
of $P(\Delta g)$ in Fig. 6 illuminating the gradual progression towards the melting.

\begin{figure}[t]
\includegraphics[width=3.3in,height=3.0in]{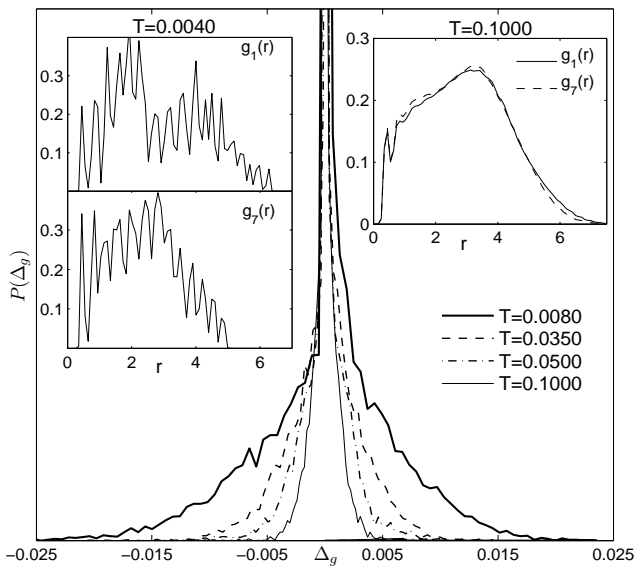}
\caption{The main panel shows the distribution $P(\Delta g)$ for $N=141$ (defined in the text). Traces of individual $g_i(r)$ for two particles ($1$ and $7$ in our nomenclature) are shown on the left-insets at low $T$, showing significant particle-to-particle fluctuations in it, which are attributed by the irregular confinement. At large $T$, on the other hand, such fluctuations vanish, producing $g_i(r)$ similar to $g(r)$ for all $i$, as shown on the right-inset for the same two particles. The resulting $P(\Delta g)$, thus becomes narrow and symmetric with $T$ in the same range $\Delta T_X$.}
\label{fig:grpar}
\end{figure}

\begin{figure}[t]
\includegraphics[width=3.3in,height=3.0in]{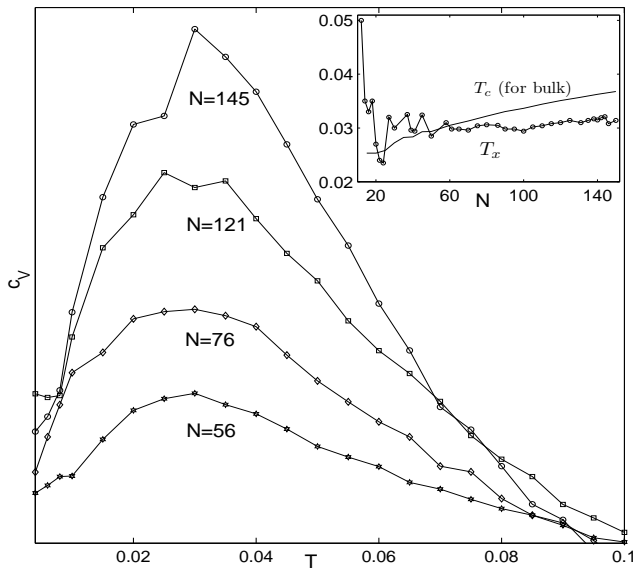}
\caption{The specific heat, $c_V$, as calculated from energy fluctuations in MC (see text), showing a hump characterizing a crossover in critical phenomena. The hump gets sharper as $N$ increases. Also, it is interesting to note that the hump occurs at $T_X=0.03$, and is insensitive to $N$. The inset compares the $N$-dependence of $T_X$ (obtained from the location of the hump in $c_V$) and $T_c$, the transition temperature for a bulk system with the same average density of particles, as in our irregularly confined system for each $N$. Modulo the large uncertainty in $T_X$ (Note $\Delta T_X \sim T_X$), their behavior indicates that the thermal fluctuations could destabilize an IWM more than an equivalent bulk system.}
\label{fig:SphHeat}
\end{figure}

Having seen a consistent $\Delta T_{X}$ for solid-like to liquid-like
behavior in an IWM, we turn to the $T$-dependence specific heat, $c_{V}$, of
the IWM,\cite{SpeHeat08}. With $E=E_{MC}$ -- the total MC energy, and $\hat E=\langle E \rangle$, $c_{V}$ is defined as,
\begin{equation}
c_{V} = \frac {d\hat E}{dT} = (\langle E^2 \rangle - \langle E \rangle ^2) / T^2.
\end{equation}
The temperature evolution of $c_{V}$
is presented in Fig. 7, showing a distinct hump as expected in a CO, which gets sharper as $N$ increases. It is important to
note that the position of the hump, $T_{X}\approx0.03$, is fairly
insensitive to $N$, and falls close to the midway of $\Delta T_{X}$. We also emphasize that it is the same $T_X$, at which the distribution $P({\cal L}_i)$ developed a peak at a non-zero value of ${\cal L}_i$ for the first time as $T$ was increased from zero (See top left inset of Fig. 2).
The estimate of the $T_{X}$ from the peak-value of smoothed $c_V$ is presented in the inset of Fig. 7 as a function of $N$. We also present for comparison the corresponding $T_c$ for bulk using
 $\sqrt{\pi \overline{n}}/T_c \approx 137$ from Ref. \onlinecite{GannSudip79}, where the value of average density, $\overline{n}$, is obtained from our results on IWM. Interestingly, $T_X$ is found to lie lower than the $T_c$ for $N \geq 35$.
This is qualitatively consistent with what happens in a circular Wigner molecule \cite{BedPeet94}. However, serious significance might not be associated with this comparison, because the uncertainty in $T_X$ is large ($\Delta T_X \sim T_X$).
Based on different criteria for $T_{X}$ and more broadly for
$\Delta T_{X}$ that we report here, we find that the melting is strongly smeared, with the width of transition comparable to the melting temperature.

\section{Mechanism for the Crossover}\label{sec:Mech}

Our results evidently raise the next fundamental question: What's the
mechanism driving the crossover found in our IWM? This is particularly important in
comparison with the established mechanism of melting in the bulk 2D system, and circularly confined systems.
As discussed before, the thermal melting of 2D WC, described by KTHNY theory, is a two-step process mediated by production of crystal defects, e.g., dislocations and disclinations leading to the breaking of positional and orientation orders respectively. Melting in circular confinements, also a two-step melting process, is enforced by the symmetry, delocalizing particles along the azimuthal direction in one step and melting along radial direction in the other. As confirmed already -- radial and azimuthal melting loose relevance in our $V_c$ due to its complete lack of symmetry. Can we still identify any crystal defects? Can their thermal evolution help in understanding the CO encountered?

In absence of a firm analytical theory describing `melting' in irregular confinements, we turn to numerical evidences from our calculations and we find the following: The positional order is largely depleted in the IWM even at the lowest temperature, and this is consistent with our occasional finding of signatures of dislocations, that resemble an extra row of particles stuck partly, in the solid-like phase. However, we note that rigorous identification of dislocations is difficult due to: (a) smallness of the system itself, and (b) presence of irregularities resulting into inhomogeneous and irregular lattice structures. An example of a dislocation found in a realization of the confinement is shown in Fig. 8(a). The thermal evolution of such dislocations is less clear for the reasons above, but to the extent we can infer, we do not see any trend for $T$-dependent proliferation of them.

Disclinations, on the other hand, characterized by a mismatch in the orientation as one circumnavigates it, signal the loss of BOO and are indeed found to proliferate for all $T\geq0.01$ in our data. They are best seen as a particle having the ``incorrect" number of nearest neighbors as we discuss below.
The number of nearest neighbors of a particle, also called the coordination number (CN) of that particle, in a given configuration is best measured using the Voronoi diagram (VD) \cite{voronoi88}. The VD assigns a 2D polygon around each particle in a given configuration, such that, any point within that polygon will be closer to that given particle than from all other particles. Thus, the VD determines the coordination number for all the $N$ particles, and identifies their closest neighbors in an unbiased manner. Obviously, the VD corresponding to a perfect triangular lattice would be regular hexagons of same size touching each other with lattice points lying at the centers, and each particle on such a lattice will have $6$ nearest neighbors ($CN=6$).

\begin{figure}[t]
\includegraphics[width=3.3in,height=3.3in]{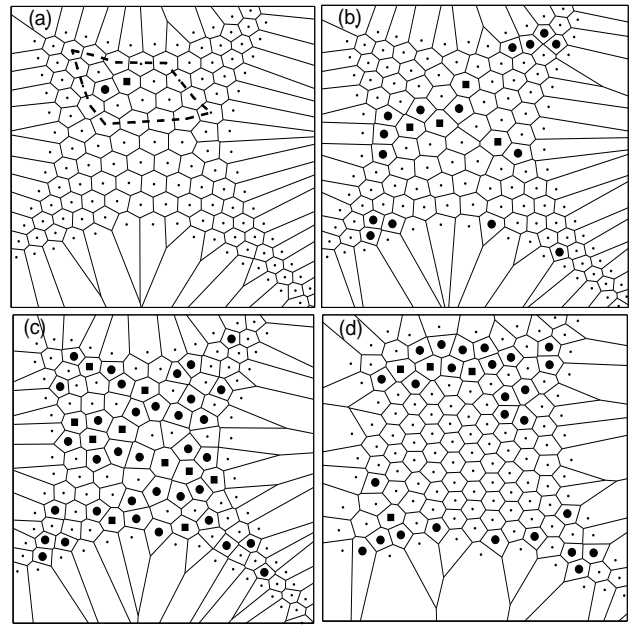}
\caption{ The Voronoi diagram corresponding to single equilibrium MC configuration with $N=148$ corresponding to Fig. (1). The location of  actual particles are shown by thin dots for those with $CN=6$, filled squares for $CN=7$, and filled circles for $CN=5$. Note that $CN=5,7$ implies local disclinations.
(a) VD for GSC at $T=0$ shows predominantly regular hexagonal area surrounding each particle, resulting strong BOO. Note the identification of a dislocation even at $T=0$ (as discussed in text) by the region bounded by thin dashed line that contains a bound pair of disclinations, as expected. (b) The configuration at $T=0.015$ illustrates the beginning of the formation of a correlated path of free disclinations. (c) The proliferation of disclinations is demonstrated on the VD at $T=0.065$ (d) The VD for hard-core particles implies that disclinations are limited primarily near the boundary leaving sizable BOO in the bulk.}
\label{fig:Defect}
\end{figure}

We present three VDs on an arbitrary equilibrium MC configuration particles in Fig. 8 for the same realization of confinement as in Fig. 1 (a-c). The result at $T=0$ shows CN$=6$ for nearly all particles, except for those on the boundary and for the bound pair of disclinations contained within the dislocation. The VDs at $T=0.015$, and $T=0.065$ illustrate progressive proliferation of the disclinations (i.e., particles with CN$=5$ or $7$) with $T$.

\begin{figure}[t]
\includegraphics[width=3.3in,height=4.8in]{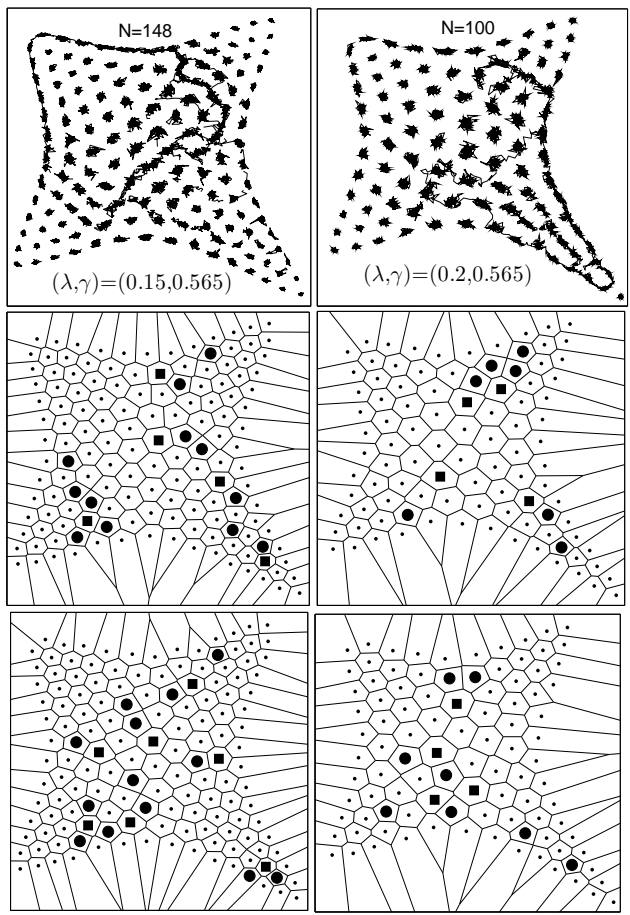}
\caption{The configuration for two different realizations of ($\lambda,\gamma $) with $N=148$ and the Voronoi diagrams corresponding to two different equilibrium MC configurations. Filled squares for CN $=7$ and filled circles for CN$=5$. The disclination mainly has started along the line of the tortuous paths and also the Voronoi plaquettes corresponding to the particles along such line is rather distorted than those of other particles.}
\label{fig:Disclination}
\end{figure}

We now come back to the question: Why does the Lindemann ratio show rapid increase for $T\geq0.01$? Analysis of our numerical data indicates that for $T\geq0.01$ disclinations start growing causing re-adjustments of particles in space, which expectedly raises ${\cal L}$. Such re-adjustments obviously affect the particles close by, leading naturally to spatially correlated movements of particles in certain regions in space. This not only explains the incipient melting through the tortuous paths found in Fig. 1b, but also the large value of $\sigma_{{\cal L}}$ (Fig. 2) for $T$ in the range of $\Delta T_{X}$ as well. Such a physical picture for melting was found consistent with our data obtained for $N\geq 40$, and also across all the realizations of confinement.

In order to emphasize the last point, we present the evolution of the disclinations for two other realizations of the confinement in Fig. 9. The top panels show the snapshot of particles over $100$ independent MC steps at $T=0.02$, where the melting has just commenced. These are thus similar to Fig. 1(b), but for different realizations, as well as for different $N$. We show the corresponding VDs for two profiles for independent MC steps in the middle and bottom panels. These figures demonstrate that the tortuous path of the melting is rife with disclinations. We further observe that the particles with CN$=6$ close to such tortuous paths are more likely to have distorted hexagons as the Voronoi-plaquettes surrounding them than the others away from such paths. Our results, therefore, suggest strongly that the crossover in an IWM from solid-like to a liquid-like phase is associated with disclinations destroying BOO. The emergence of such a mechanism constitute the key finding of our study.

\section{Discussion}\label{sec:discus}

We found that the IWMs with $N\geq35$, present qualitatively similar physics discussed so far. But, for smaller values of $N$, the crossover from a solid-like to a liquid like behavior is lot more smeared, and is rife with larger fluctuations in all physical observables. Further,a higher ratio of boundary to bulk particles being for smaller $N$ makes it harder to look for `universal' features.

All the results discussed here are for $a=1$ in $V_c$ (Eq. 1), which essentially fixes the average density of particles in the system, for example, $\overline{n}=6.857$ for $N=100$. We have repeated the same calculations for lower densities on a few realizations, namely for $a=0.1, 0.01$ resulting into $\overline{n}=2.73$ and $\overline{n}=1.06$, respectively for $N=100$, and find qualitatively similar results. However, $T_X$, as well as $\Delta T_X$ were found to decrease with $\overline{n}$. In any case, such inferences are expected to change at very low densities where quantum effects become significant, particularly at lowest temperatures. Those effects, while significant, are beyond the scope of the current work and remain as an important future direction.

In conclusion, we have studied the thermal crossover from a Wigner-type
solid-like to a liquid-like phase in an irregular confinement containing
classical particles. Our results demonstrate that such crossover
takes place gradually without any sharp changes, and hence the width
$\Delta T_{X}$ is large. Thermal evolution of different observables
points towards a unique $\Delta T_{X}$. Interestingly, the mechanism
for melting appears to be the proliferation of disclinations that
destroys the quasi-long range orientational order. Breaking of all symmetries
does not, as such, stabilize the quasi-order state in IWM than in the bulk. Experiments study some of the dynamical properties of melting, such as, diffusion constant, frequency dependence of structure factor, the dynamics of the defects
and their role in melting  etc. While our method constrains
us to study only the static properties, an extension to include the dynamics of the defects seems to be a bright direction. We hope that our finding
will help understanding the physics of chaotic quantum dots in the
experimental regime.

\begin{acknowledgments}
 We would like to thank D. Dhar, S. Lal, H. U. Baranger, D. Sen and J. Chakrabarti for valuable conversations.
\end{acknowledgments}

\bibliography{qdotqmc,footnotes}

\end{document}